# Metastable behavior of Urbach tail states in BaTiO$_3$ across phase transition

Vikash Mishra[1], Archna Sagdeo[2,3], Kamal Warshi[1], Hari Mohan Rai[1], Shailendra K. Saxena[1], Rajesh Kumar[1] and P.R. Sagdeo[1]*

[1]*Material Research Laboratory, Discipline of Physics and MEMS, Indian Institute of Technology Indore-453552, India*
[2]*Indus Synchrotron Utilization Division, Raja Ramanna Center for Advanced Technology Indore-452013, India*
&
[3]*Homi Bhabha National Institute, Training School Complex, Anushakti Nagar, Mumbai-400094, India*

**Abstract**

The temperature dependent diffuse reflectance spectroscopy measurements were carried out on the polycrystalline samples of BaTiO$_3$ across the tetragonal to cubic structural phase transition temperature. The values of various optical parameters such as band gap ($E_g$), Urbach energy ($E_U$) and Urbach focus ($E_0$) are estimated in the range of 300 K to 470K. It is observed that near structural phase transition temperatures there exists two value of $E_0$, suggesting presence of electronic heterogeneity over wide temperature range. Further near transition temperature $E_U$ shows metastability i.e. value of $E_U$ at temperature T is not constant but is a function of time (t). Interestingly it is observed that the ratio of $E_{U(t=0)}/E_{U(t=tm)}$, is almost remains constant at 295 K (pure tetragonal phase) and at 450 K (pure cubic phase), whereas this ratio shows decreasing behavior close to structural phase transition temperature, which confirms the presence of electronic metastibility in the pure BaTiO$_3$. The observed metastibility can be fitted with the stretch exponents relaxation behavior, suggesting the presence of dynamic heterogeneous electronic disorder present in the sample across the transition. Further it appears that these metastable Urbach tail states (electronic disorder) may couple with the soft phonon modes and responsible for the observed terahertz dielectric relaxation (Phys. Rev. Lett. **101**, 167402 (2008)). Further; present studies suggest that the optical studies appear to be more sensitive to probe the disorder/heterogeneity present in the sample.

*Corresponding author: **prs@iiti.ac.in**

## Introduction:

Understanding the variation of $E_g$ as a function of temperature and that of $E_U$[1,2] are of scientific and technological interests[3,4]. It is well accepted that the $E_U$ which is a measure of various disorders present in the system (thermal, polar, chemical, structural and due to defects etc.) should be small enough for pure un-doped system. The various contributions to the $E_U$ and their temperature dependence is described and discussed in literatures earlier[5]. Generally it is observed and believed that that the $E_U$ scales up with thermal disorder i.e. increases with the temperature. For the samples which undergo the phase transition an anomaly is observed in the temperature dependent behavior of $E_U$. Zametin[6] in 1984 presented the detail review and understanding on the anomalies in optical properties of polar materials at phase transitions and shown that; in case of polar materials such as polar semiconductors and dielectric materials; there exists huge change in electron-phonon interaction which results in the change of the shape of the absorption edge at transition temperature, further it is pointed out that scattering of carriers and perturbation of their wave functions also results in the change of the half-width and amplitude of the absorption[6] and the uniform statistical distribution of two or more phases is considered near to the transition temperature[7]. In past the efforts have been made to understand the temperature dependent mechanism(s) underlying the Urbach rule[8,9]. The exponential behavior of Urbach energy as originally described by Urbach is of the following form.

$$S = \frac{\partial (\ln K)}{\partial (\hbar \omega)} = -\frac{1}{kT} \qquad (i)$$

Here α is absorption coefficient, E is the photon energy, $k$ is the Boltzmann constant and T is temperature. The above equation suggests that S is independent of material and completely governed by temperature, but experimentally it is observed that the S is purely a property of material and is a function of temperature, hence Martienssen[10] redefine the Urbach rule and wrote the same in the following form.

$$\alpha(E,T) = \alpha_0 \exp[\sigma(\frac{E - E_0}{kT})] \qquad (ii)$$

Here $\alpha_0$ and $E_0$ are the constants determined by extrapolated linearly from $ln(\alpha)$ versus E curves at a given temperature T. The functional form of σ(T) is given by Mahr[11] and is as follows

$$\sigma(T) = \sigma_0(\frac{2kT}{h\nu_p}) \tanh(\frac{h\nu_p}{2kT}), \qquad (iii)$$

Here $\sigma_0$ is a constant according and found to be inversely proportional to electron phonon interaction; and $\hbar\omega_0$ the energy of phonon most strongly bound with the electron[6]. While discussing above stated Urbach rule at the phase transition it is pointed out that the Urbach rule[22] as discussed in equation No.(ii) appears to be more correct and one need to take in to account the anomaly in Eg and change in the shape of absorption edge. The above mentioned literatures clearly suggest and confirm that the shape of Urbach tail states carries the information about the phase transition[12–14].

BaTiO$_3$ is one of the most studied classical and an important ferroeletric material and known to show the first order structural phase transition[15–18]. Further it is expected that at the structural phase transition BaTiO$_3$ may show change in electron-phonon interaction which may results in the change of the shape and amplitude of the absorptionedge.[6]

Even though the temperature dependent optical studies[19,20] are available on BaTiO$_3$[21–23] but no clear understanding about the behavior of temperature dependence of Urbach tail states are available[24]. Further BaTiO$_3$ is known to show relaxation in terahertz region[25]. The origin of such relaxation is proposed to be is partially soft S-phonon mode and for order disorder dynamics associated with stochastic interwell hops[25]. It is important here to note that the corresponding equivalent frequency of these Urbach tails states (Urbach energy) for Barium Titanate is in THz region. Further there are reports suggesting evidence of small polaronic effect (electron lattice coupling) in BaTiO$_3$[26–28]. Thus it appears that in energy band there exist finite overlaps between phononic and electronic density of states. Further there are evidences of relaxation in central Raman peak[29], thus; if electron phonon coupling exist it must produce corresponding relaxation in electronic states i.e. Urbach tail states, keeping this in view here we have performed temperature dependent diffuse reflectance spectroscopy and report the temperature dependent band gap, Urbach energy, stiffness coefficient and Urbach focus across phase transition. It is observed that near phase transition there exists two Urbach focus on corresponding to tetragonal phase and cubic phase, thus it appears that the optical band gap, Urbach energy, steepness coefficient etc. may have finite contribution from both the crystallographic phases over wide temperature range across the phase transition. Further it is observed that the Urbach energy which is measure of various disorder present in the sample shows metastability across the phase transition and the observed metastibility can be explained on the basis stretch exponents relaxation behavior, suggesting the presence of dynamic heterogeneity present in the sample across the transition temperature. The present experimental results clearly demonstrate that the Urbach tail states carry the information about the various disorders present in the system including that of structural one (random distribution of tetragonal and cubic), further it these metastable states show hysteresis in wide temperature range as compared to that of band gap and it appears that these electronic states due to various disorders couples with the phonon and may produce relaxation in terahertz frequency range.

**2. Experimental Details:**

(i) Sample Preparation: Polycrystalline samples of BaTiO$_3$ were prepared by the conventional solid-state reaction route using BaCO$_3$ (99.99%) and TiO$_2$ (99.99%)[30].

(ii) **Structural Characterizations:** In order to examine the structural phase purity of the prepared samples the powder x-ray diffraction (XRD) experiments were carried out on Bruker D8 diffractometer equipped with Cu target. The high temperature x-ray diffraction measurements were performed to confirm the structural phase transition in the prepared samples.

(iii) **Dielectric Measurements:** Temperature dependent dielectric measurements were carried out using a WayneKerr precision impedance analyzer [31,32].

(iv) **Temperature dependent Diffuse Reflectance Measurements:** The optical band gap of prepared samples has been measured using diffuse reflectivity measurements. These measurements have been performed in the 200 nm to 800 nm wavelength range using Cary-60 UV-VIS-NIR spectrophotometer having Harrick Video-Barrelino diffuse reflectance probe in the temperature range of 300K to 480K.

**Results and discussion**

   (I)     **Structural characterization and phase purity.**

Figure-1 shows the powder x-ray diffraction pattern for the prepared sample. The obtained x-ray diffraction data is indexed considering the space group P$_4$mm and no impurity peaks are observed, which confirms the structural phase purity of the prepared sample. The high temperature x-ray diffraction measurements confirm the structural phase transition at around ~ 398 K as shown in the inset of figure-1.

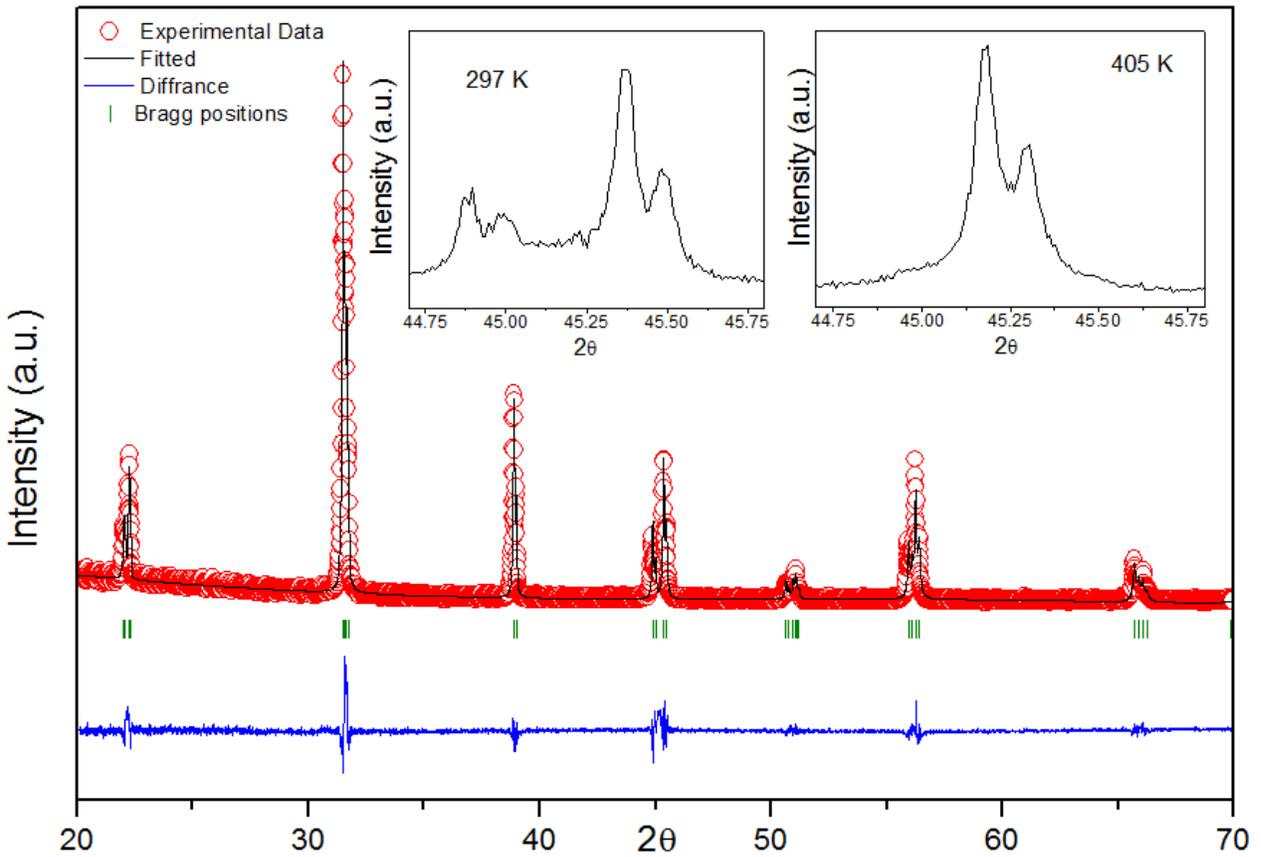

*Figure 1: Powder x-ray diffraction pattern for prepared BaTiO₃ the absence of any unaccounted peak confirms the phase purity of the sample. The inset shows the (002) at room temperature and at 405K.*

(II) **Dielectric Measurements:** Figure-2 shows the temperature dependent dielectric constant for the prpared $BaTiO_3$ sample at various frequencies. The signature of the ferroelectric to paraelectric transition (as governed by tetragonal to cubic structural phase transition) is clearly visible. The dielectric data confirms the absence of dielectric relaxation in the prepared $BaTiO_3$ sample[30]. These high purity samples were used for the optical characterization.

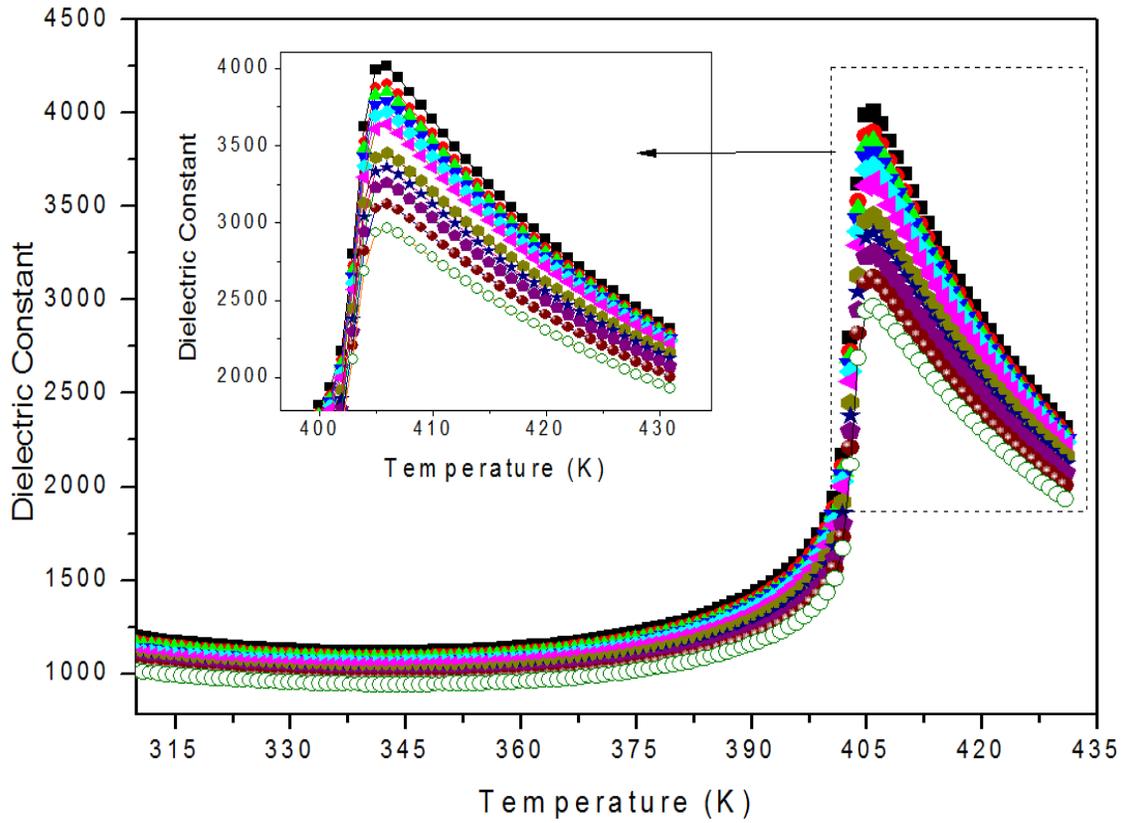

*Figure 2 Temperature dependent dielectric data for BaTiO$_3$. The inset shows the magnified view around transition temperature confirming the absence of dielectric relaxation in the prpared sample in the studied frequency range.*

### (III) Diffuse Reflectance Spectroscopy (DRS)

In the present study we have probed the optical absorption for BaTiO$_3$ using diffuse reflectance spectroscopy, the spectra obtain from DRS is converted into equivalent absorption spectra through Kubelka–Munk[33,34] equation.

$$F(R_\infty) = \frac{(1-R_\infty)^2}{2R_\infty} \qquad \text{(iv)}$$

Where $F(R_\infty)$ is the Kubelka–Munk function, $R_\infty = R_{sample}/R_{standard}$, $R_{Sample}$ is the diffuse reflectance of the sample and $R_{Standard}$ is that of the standard (BaSO$_4$ in present case). Figure 3(a) shows graph between obtained Kubelka–Munk function versus wavelength for prepared BaTiO$_3$ sample at various temperatures. Keeping in view the perfect diffuse scattering from the prepared sample the Kubelka–Munk function can be related (proportional) to the absorption coefficient ($\alpha$) as[35]

$$F(R_\infty) \propto \alpha \propto \frac{(h\nu - E_g)^{1/n}}{h\nu} \qquad \text{(v)}$$

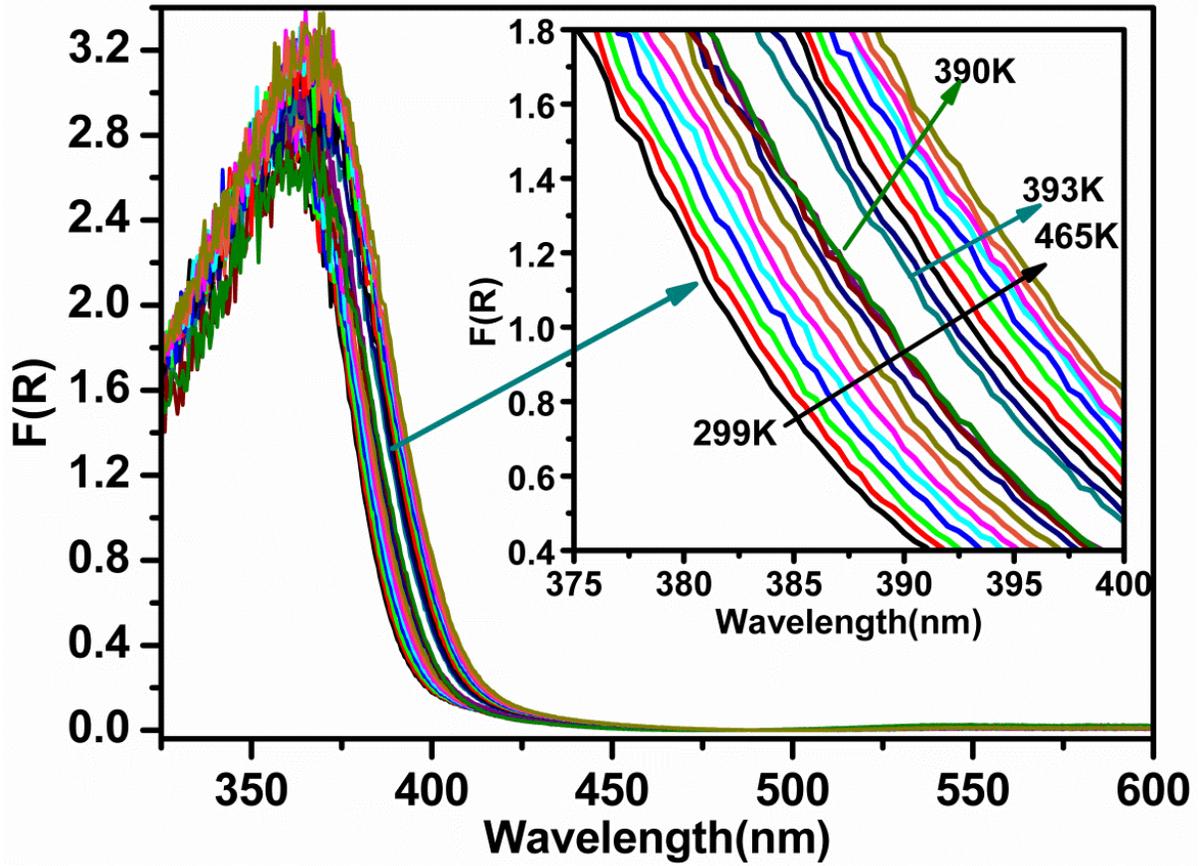

*Figure 3(a) Plot between kubelka-munk function and wavelength of BaTiO₃ sample at various temperatures (299K-465K).*

In order to calculate the Eg the obtain absorption coefficient is converted in to Tauc equation[2] and plotted in figure 3(b).

$$(\alpha h\nu)^n = A(h\nu - E_g) \qquad (vi)$$

Here in equation(vi) n has the value of 2 for direct bandgap transitions, while n is equal to 1/2 for an indirect transition[35]. We have used the values of n = 2 to determine the optical gap of BaTiO$_3$ as is proposed as direct band gap material[21], and known to show photoluminescence properties[36].

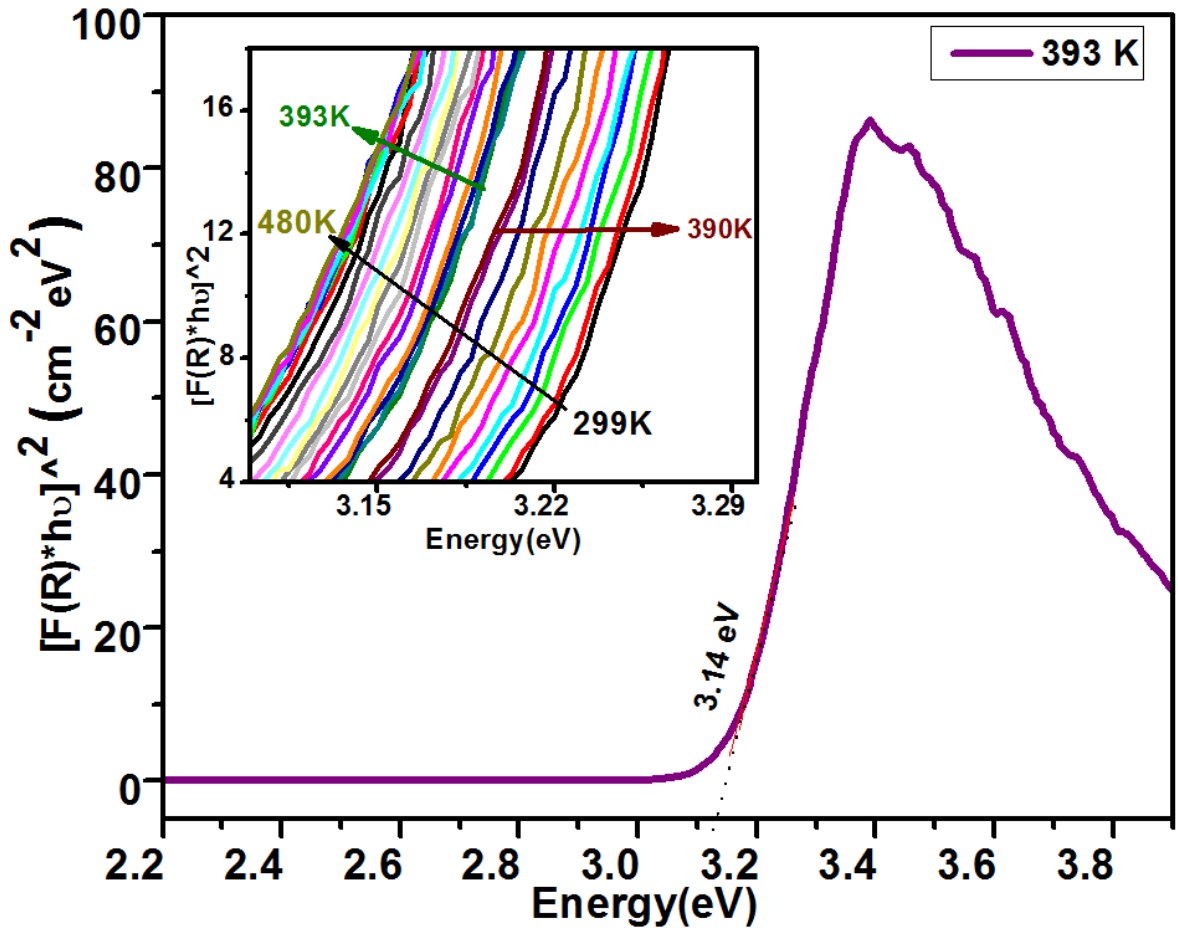

*Figure 3(b):- Tauc plot to determine the Band gap of BaTiO₃ at various temperature.*

Figure 4(a) shows the variation of band gap as a function of temperature. From the figure 4(a) it is clear that with increase in the temperature the band gap of BaTiO$_3$ systematically decreases and there exists an anomaly at ~395 K which is a well know tetragonal to cubic structural phase transition temperature for BaTiO$_3$[37]. It should be noted that above and below structural phase transition it is possible to verify the variation of band gap as a function of temperature using Varshni's relation[38–40] as shown in the inset of figure 4(a). Figure 4(b) shows the heating and cooling plots of band gap Eg for the studied samples, suggesting very narrow hysteresis in band gap the studied temperature range. It should be noted that BaTiO$_3$ shows the first order structural phase transition[37]. Thus it is expected that near to the phase transition there exists structural and polar disorder[3,5], Thus it is expected that Urbach tail states i.e. Urbach energy should have the contribution due to various above mentioned disorders along with the thermal one. Hence we have critically looked into the variation of Eu as a function of temperature near phase transition, the details of the same are provided below.

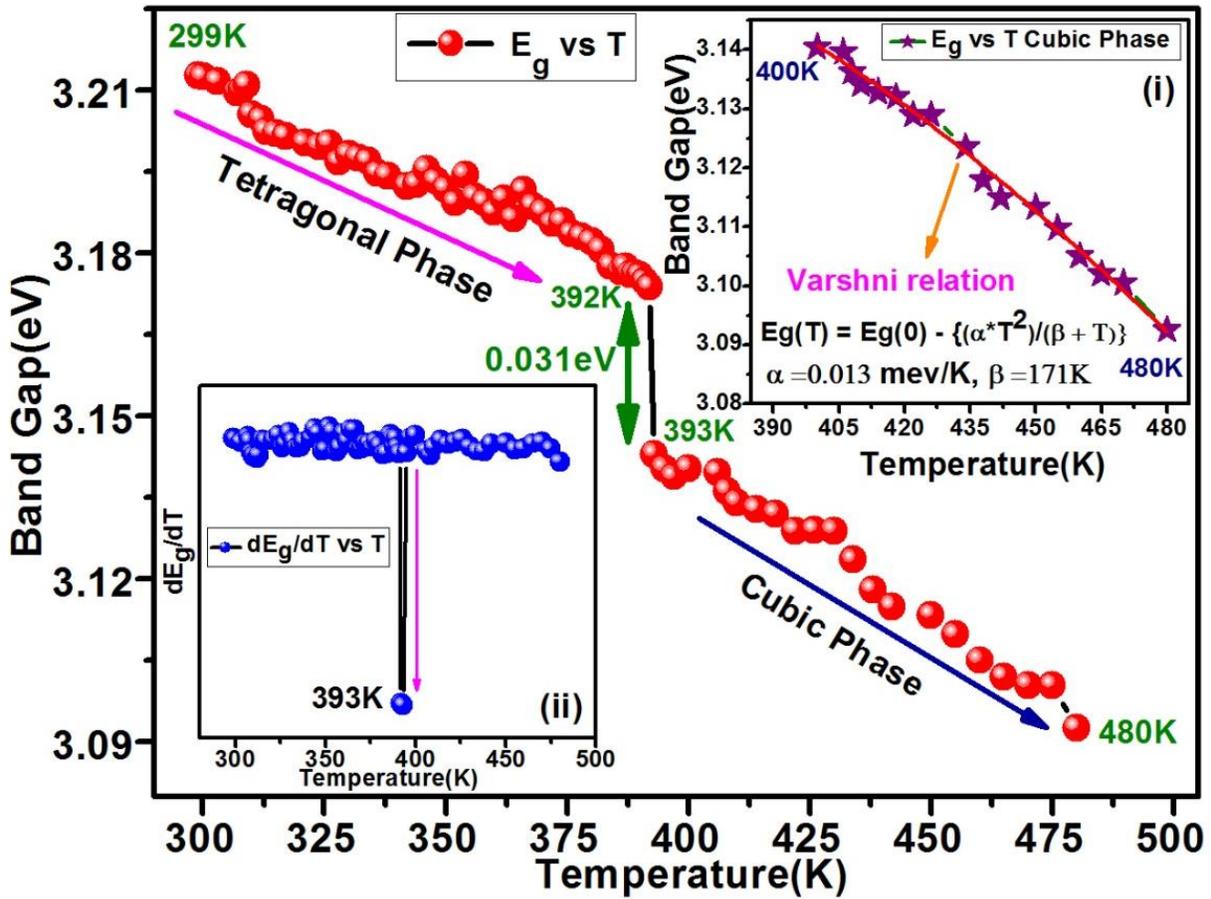

*Figure 4(a):- Variation in Band Gap ($E_g$) at various Temperature (299K-480K) of BaTiO₃. Inset (i) shows fitting of cubic phase by using theoretical model (varshni's relation) inset (ii) shows plot of $dE_g/dT$ to determine transition temperature.*

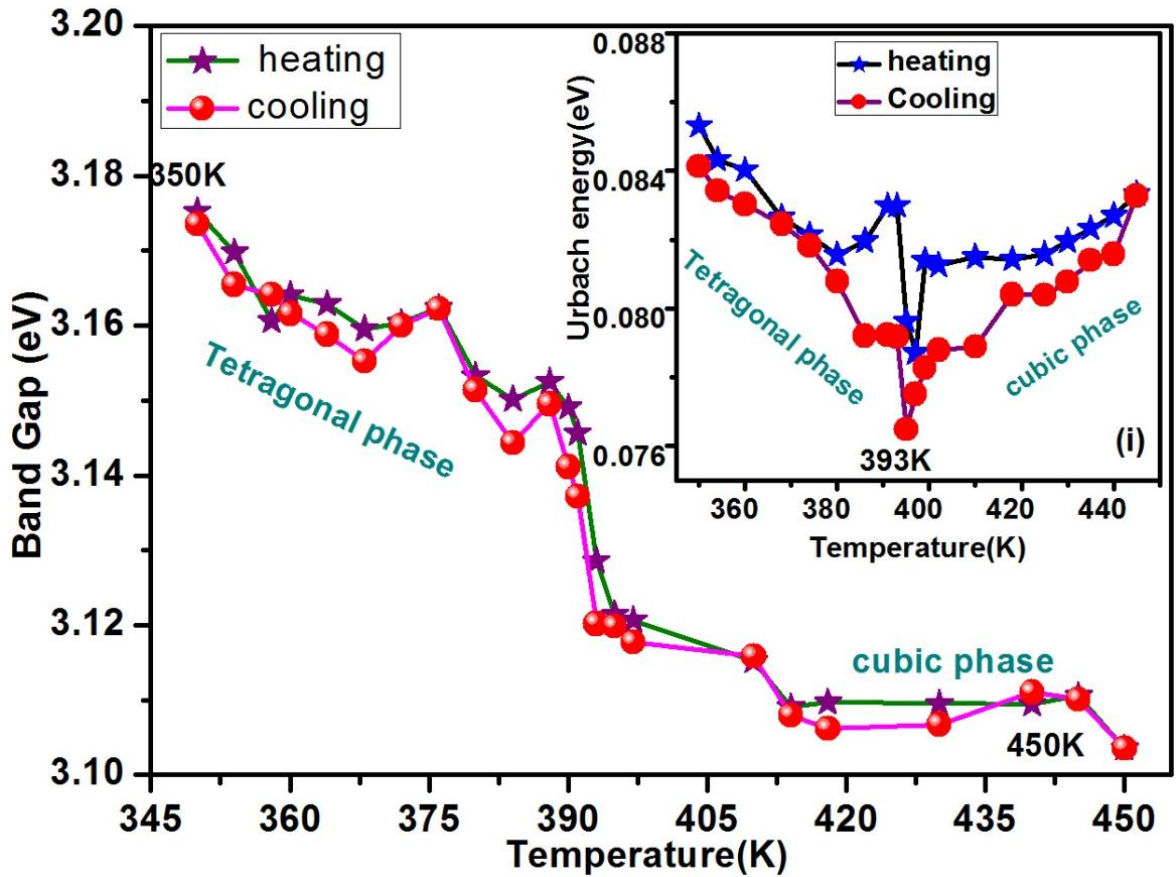

*Figure 4(b): Variation in band gap during heating and cooling cycles which confirms reversibility value of band gap of BaTiO$_3$ inset (i) shows the corresponding hysteresis in E$_U$.*

**Urbach energy and Urbach Focus:** As discussed above the Urbach energy provides the information about the disorder present in the sample. The contributions to the said energy may be due to various types of disorders such as thermal, polar, chemical, structural and due to defects etc.[5]. The inset of figure-4 (b) shows the variation of E$_U$ as a function of temperature. At the tetragonal to cubic structural phase transition E$_U$ suddenly decreases, this may be due to the disappearance of some of the vibration mode in Raman spectra[4,41,42] of cubic phase in BaTiO$_3$ leads to the sharp decrease in the overall value of Urbach energy at transition temperature further cubic phase being more order phase as compared to that of tetragonal phase may also lead to decrease in E$_U$.

In order to further investigate the effect of phase coexistence (structural and polar heterogeneity) on the optical properties we have plotted natural log of absorption coefficient α for BaTiO$_3$ as a function of incident photon energy[5,43] as shown in figure 5(a) and 5(b). Two Urbach focuses[5] which is a disorder-independent constant which seems to be related to the band-edge[43,44] corresponding to cubic and tetragonal phases are clearly evident for high temperature and low temperature data. Surprisingly in the temperature region of 384K-480K two slopes are clearly visible as shown in figure 5(c). From this figure it is clear that it is possible to extract two Urbach focuses whose values are very close to that of cubic and tetragonal phases. Our results are consistent with those reported by Park et al[45].

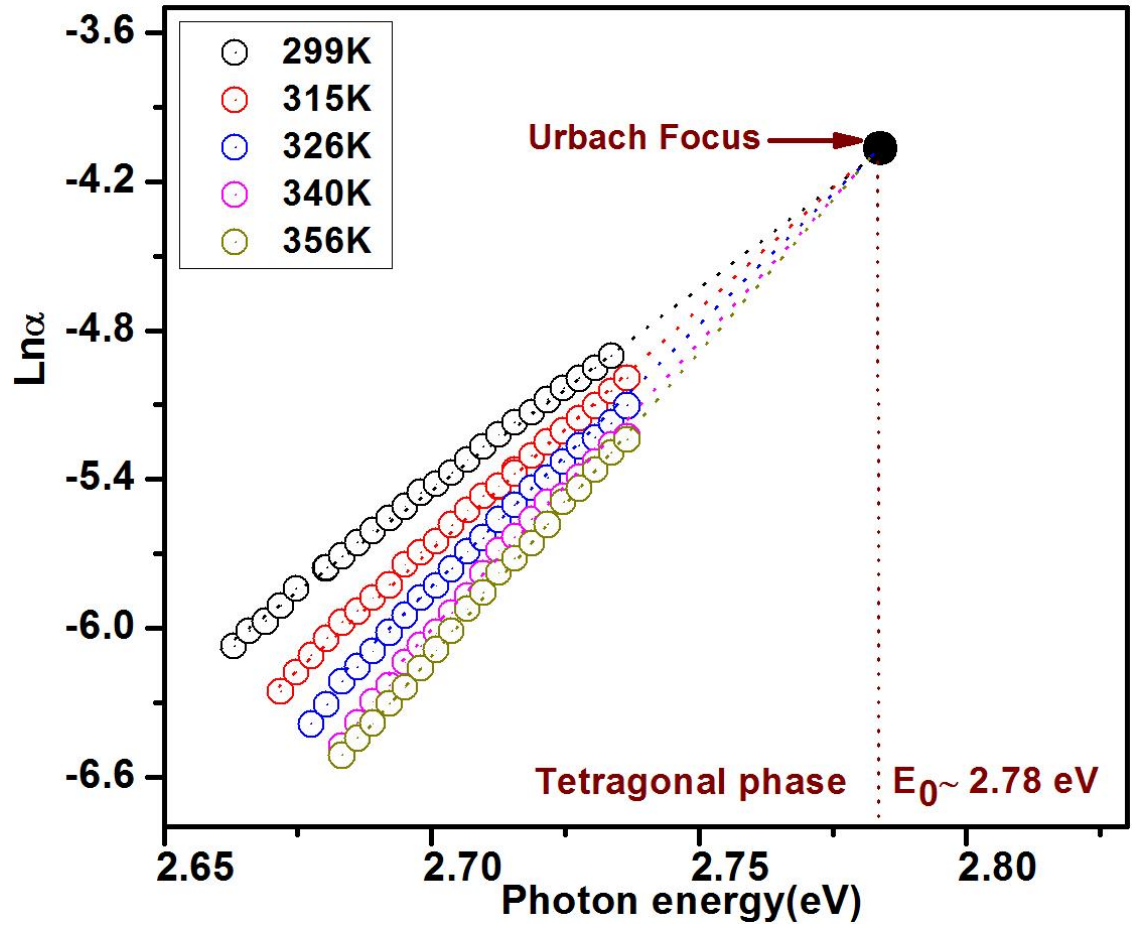

**Figure 5(a)** Urbach focus across tetragonal phase of BaTiO$_3$ for tetragonal phase.

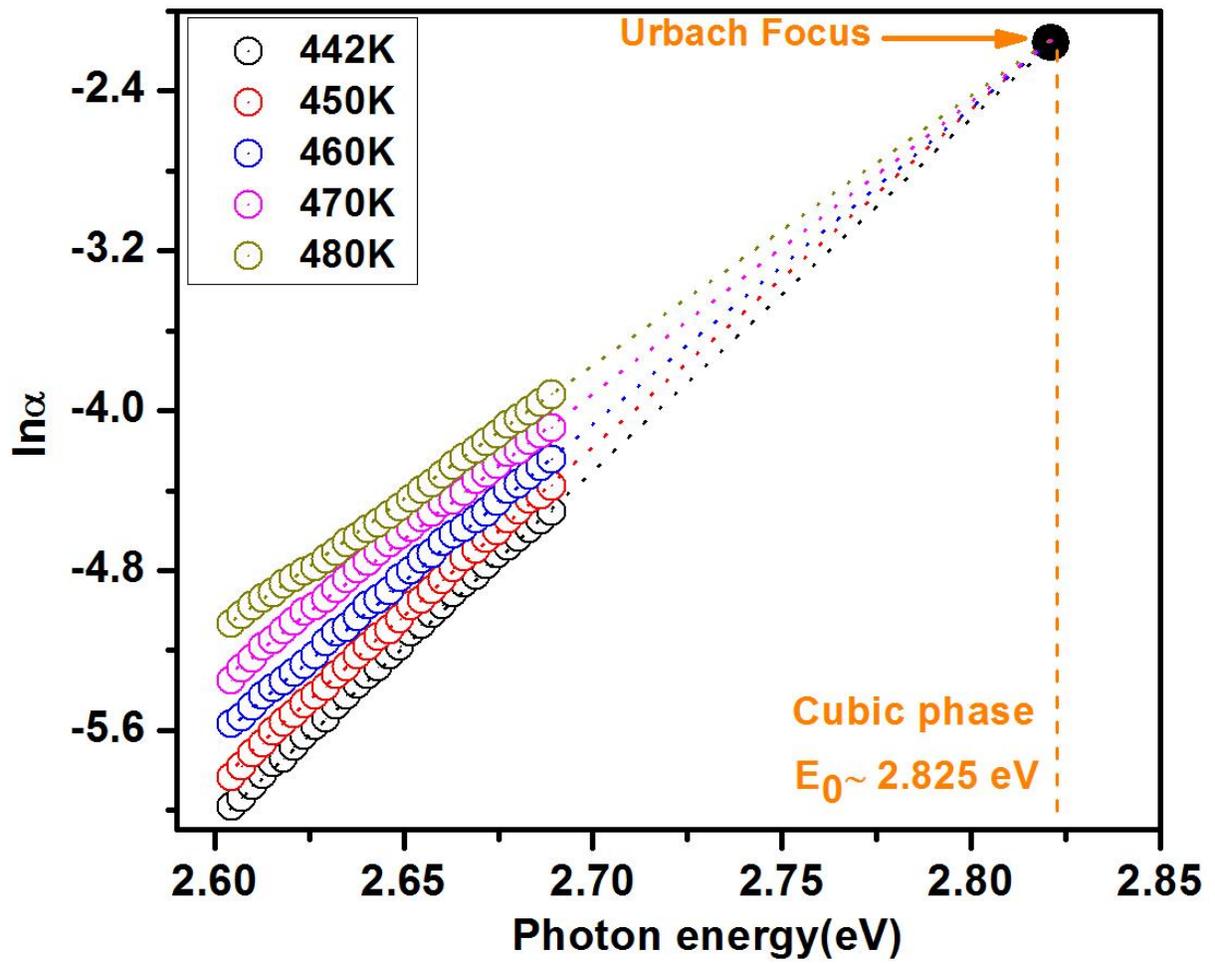

*Figure 5(b) Urbach focus across cubic phase of BaTiO$_3$ for cubic phase.*

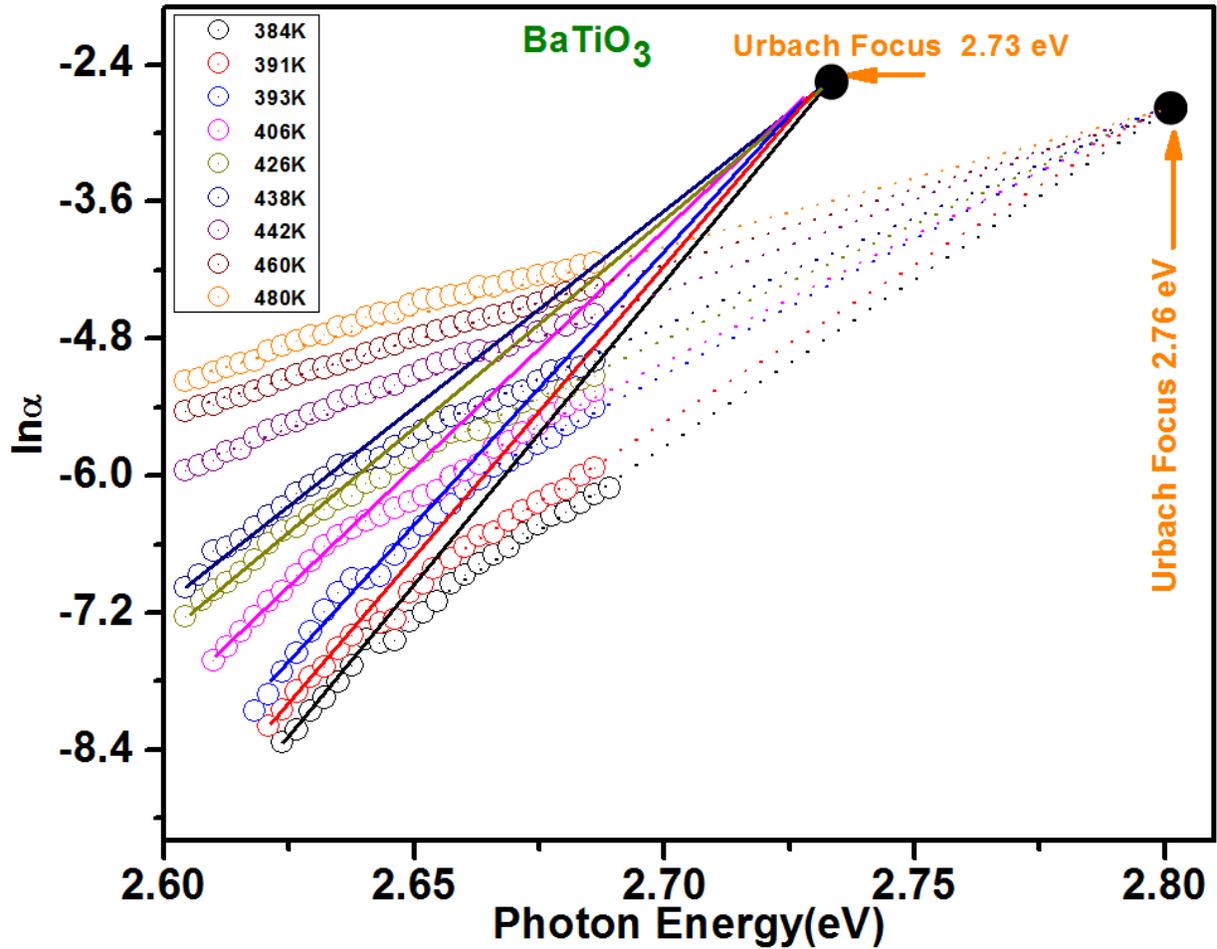

*Figure 5(c) shows two slopes for lnα versus photon energy data. The two slopes corresponds to two different Urbach focus one for cubic and other for tetragonal phase.*

In order to confirm the structural phase coexistence (structural heterogeneity) in the temperature range corresponding to two slopes the temperature dependent x-ray diffraction data was analyzed. Figure-6 shows the x-ray diffraction data at 405 K and 415 K. From the figure it is clear that with increase in the temperature the intensity of asymmetric feature as indicated by arrows decreases and intensity of (200) peak increases. This suggests that there exists the structural phase coexistence over wide temperature range.

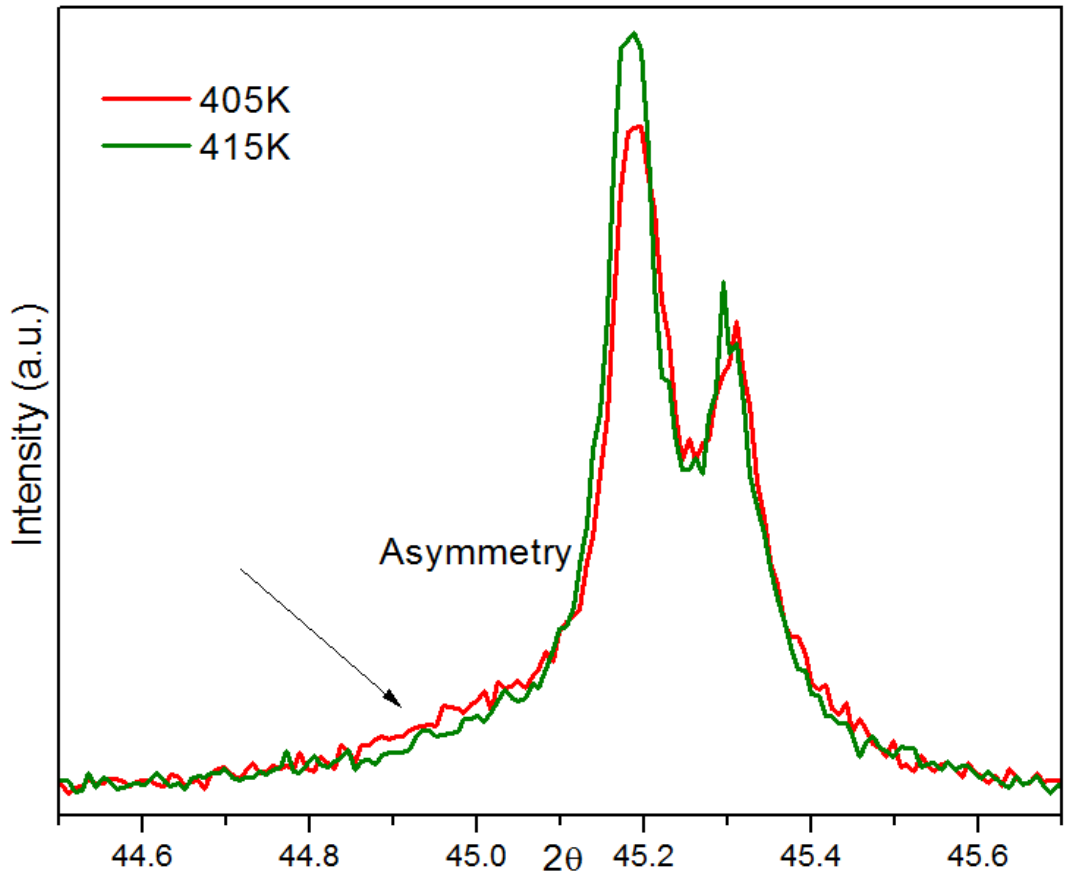

*Figure-6: High temperature x-ray diffraction data of BaTiO$_3$. The decrease in the intensity of asymmetric feature and increase in the intensity at 415 K as compared to that of 405 K is clearly visible, confirming structural phase coexistence.*

In order to further confirm the heterogeneity near phase transition we have studied the kinetics and metastability of E$_U$ around the phase transition[46–50]. In Figure-7 we have plotted the ratio of E$_{U0}$/E$_{Ut}$ at various temperatures as a function of time. From the figure it is clear that the ratio E$_{U0}$/E$_{Ut}$ is almost one in pure tetragonal (295 K) and cubic (450K) phases whereas this ratio shows systemic decrease with time near (above and below) transition temperature, this suggests that across the phase transition temperature in BaTiO$_3$ there exists multiple states/minima[4,41,51], i.e the presence of dynamic heterogeneity across the transition temperature. In order to understand the physics and kinetics of Urbach energy across the phase transition, we have analyzed the data in the terms of stretch exponents and fitted the observed data with equation (viii).

$$\phi(t) = A\exp[-(\frac{t}{\tau})^\beta] \qquad\qquad \text{(vii)}$$

The values of τ and β are provided in figure 8 which are in agreement with the expected values of τ and β[46].

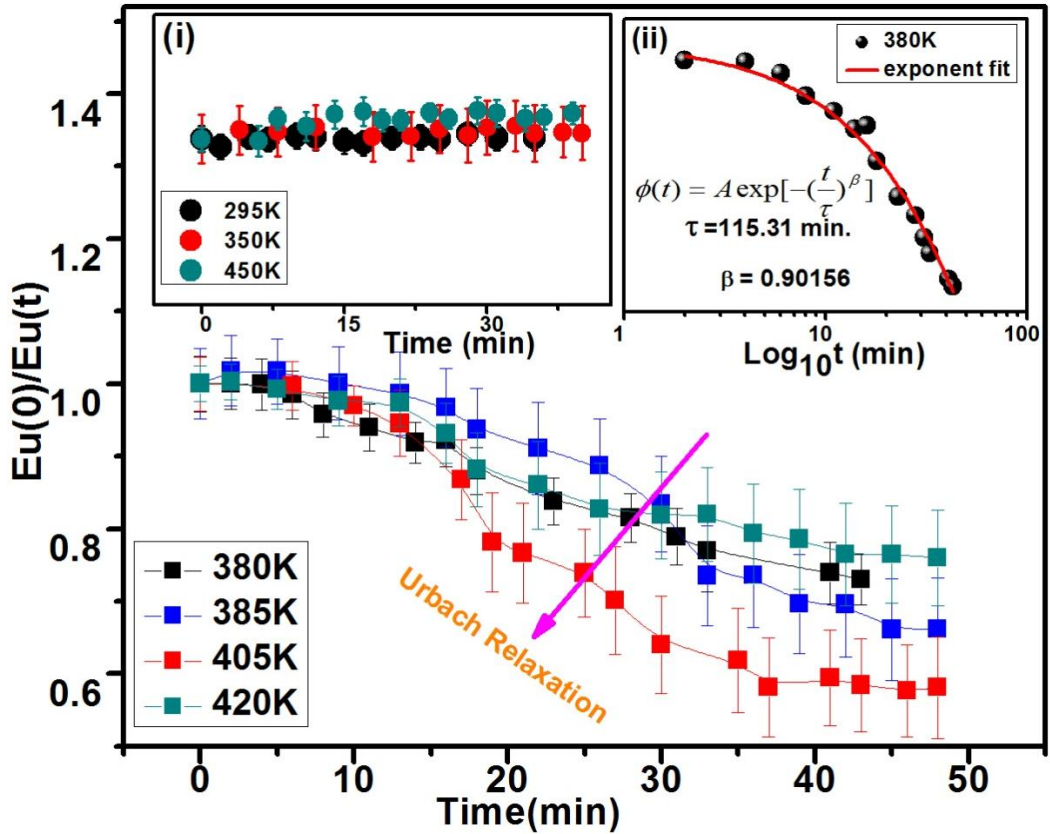

*Figure-7: plot between Eu(0)/Eu(t) as a function of time. Near transition temperature this ratio follows stretch exponent behavior. Inset (i) shows almost no variation in Eu(0)/Eu(t) in pure tetragonal and pure cubic phase. Inset(ii) Fitting of Urbach relaxation by using stretch exponents equation at 380K.*

From the above discussions it is clear that around the high temperature structural phase transition in BaTiO$_3$ there exist heterogeneity. Even though the weak signature in the form of asymmetry is present in the diffraction data but the refinement of the data may clearly miss this signature but the signature of this heterogeneity is more clearly visible in Urbach tail states, this is one of the very important observations of the present studies. At this juncture it is important to note that the dielectric measurements do not show any signature of the dielectric relaxation even when measured with very small temperature interval[35]. This suggests that the probing frequency range is not sensitive to the available disorder. J. Hlinka et. al[25].reported the coexistence of the phonon and relaxation modes in the terahertz dielectric response of BaTiO$_3$, further J. Hlinka et.al[25]. pointed out that the dielectric response around ~200-600 cm$^{-1}$ (corresponding energy of 0.025 eV to 0.075 eV) shows the dielectric relaxation. It should be noted that the energy of the Urbach tail states observed for the prepared BaTiO$_3$ sample varies from 0.076eV to 0.085eV and is very close that of terahertz range as reported by J. Hlinka et.al[25]. Thus it appears that the observed Urbach tail states may couple with the S-phonon mode[53], and may produce the dielectric relaxation in terahertz range. The present studies suggest that the optical studies appear to be more sensitive to probe the disorder/heterogeneity present in the sample.

**Conclusions:**

In summery we have shown that the pure BaTiO$_3$ shows the heterogeneity and metastability in Urbach energy in wide temperature range. It is shown that the optical spectroscopy is more sensitive to prove the disorder as compared to that of the diffraction studies and dielectric measurements. Further it appears that the observed Urbach energy states couple with S-phonon mode and produce dielectric relaxation in terahertz range. The present studies suggest that the optical studies appear to be more sensitive to probe the disorder/heterogeneity present in the sample.

**Acknowledgments**

The authors sincerely thank Prof. Pradeep Mathur, Director of IIT Indore, for his encouragement. SIC IIT Indore is acknowledged for providing some of the basic characterization facilities. CSIR, New Delhi is acknowledged for funding the high-temperature furnace under the project 03(1274)/13/EMR-II used for the sample preparations. DAE-BRNS is acknowledged for funding the impedance analyzer used for the present measurements under the grant No. 2013/37P/31/BRNS. One of the authors (VM) likes to thank Dr. Swarup Roy, Ms. Priyanka Yogi Mr. Anil Kumar and Ms. Deepika Poonia of IIT Indore for their help at various stages of this work. VM sincerely thank Ministry of Human Resource Development (MHRD), government of India, for providing financial support through teaching assistantship through IIT Indore.

**References**

[1] J.D. Dow and D. Redfield, Phys. Rev. B **5**, 594 (1972).
[2] H. Tang, F. Lévy, H. Berger, and P.E. Schmid, Phys. Rev. B **52**, 7771 (1995).
[3] A. Crisanti and L. Leuzzi, Phys. Rev. Lett. **89**, 237204 (2002).
[4] C.H. Perry and D.B. Hall, Phys. Rev. Lett. **15**, 700 (1965).
[5] M. Kranjčec, I.P. Studenyak, G.S. Kovacs, I.D. Desnica Franković, V.V. Panko, P.P. Guranich, and V.Y. Slivka, J. Phys. Chem. Solids **62**, 665 (2001).
[6] V.I. Zametin, Phys. Status Solidi B **124**, 625 (1984).
[7] M. Capizzi, A. Frova, and D. Dunn, Solid State Commun. **10**, 1165 (1972).
[8] R. Bhattacharya, R. Mondal, P. Khatua, A. Rudra, E. Kapon, S. Malzer, G. Döhler, B. Pal, and B. Bansal, Phys. Rev. Lett. **114**, 47402 (2015).
[9] F. Urbach, Phys. Rev. **92**, 1324 (1953).

[10] S.M. Wasim, C. Rincón, G. Marín, P. Bocaranda, E. Hernández, I. Bonalde, and E. Medina, Phys. Rev. B **64**, 195101 (2001).
[11] H. Mahr, Phys. Rev. **125**, 1510 (1962).
[12] K. Noba and Y. Kayanuma, Phys. Rev. B **60**, 4418 (1999).
[13] G.D. Cody, T. Tiedje, B. Abeles, B. Brooks, and Y. Goldstein, Phys. Rev. Lett. **47**, 1480 (1981).
[14] B. Abay, H.S. Güder, H. Efeoğlu, and Y.K. Yoğurtçu, J. Phys. Chem. Solids **62**, 747 (2001).
[15] J.-H. Ko, S. Kojima, T.-Y. Koo, J.H. Jung, C.J. Won, and N.J. Hur, Appl. Phys. Lett. **93**, 102905 (2008).
[16] H. Zhang, AIP Adv. **3**, 42118 (2013).
[17] K. Uchino, E. Sadanaga, and T. Hirose, J. Am. Ceram. Soc. **72**, 1555 (1989).
[18] T.-T. Fang, H.-L. Hsieh, and F.-S. Shiau, J. Am. Ceram. Soc. **76**, 1205 (1993).
[19] H.Y. Fan, Phys. Rev. **82**, 900 (1951).
[20] H.Y. Fan, Phys. Rev. **78**, 808 (1950).
[21] S.H. Wemple, Phys. Rev. B **2**, 2679 (1970).
[22] M. DiDomenico and S.H. Wemple, Phys. Rev. **166**, 565 (1968).
[23] G.A. Cox, G.G. Roberts, and R.H. Tredgold, Br. J. Appl. Phys. **17**, 743 (1966).
[24] P.K. Gogoi and D. Schmidt, Phys. Rev. B **93**, 75204 (2016).
[25] J. Hlinka, T. Ostapchuk, D. Nuzhnyy, J. Petzelt, P. Kuzel, C. Kadlec, P. Vanek, I. Ponomareva, and L. Bellaiche, Phys. Rev. Lett. **101**, 167402 (2008).
[26] P. Gerthsen, R. Groth, K.H. Hardtl, D. Heese, and H.G. Reik, Solid State Commun. **3**, 165 (1965).
[27] J.P. Boyeaux and F.M. Michel-Calendini, J. Phys. C Solid State Phys. **12**, 545 (1979).
[28] H. Ihrig, J. Phys. C Solid State Phys. **9**, 3469 (1976).
[29] V.K. Malinovsky, A.M. Pugachev, V.A. Popova, N.V. Surovtsev, and S. Kojima, Ferroelectrics **443**, 124 (2013).
[30] S. Anwar, P.R. Sagdeo, and N.P. Lalla, J. Phys. Condens. Matter **18**, 3455 (2006).
[31] H.M. Rai, S.K. Saxena, R. Late, V. Mishra, P. Rajput, A. Sagdeo, R. Kumar, and P.R. Sagdeo, RSC Adv. **6**, 26621 (2016).
[32] H.M. Rai, R. Late, S.K. Saxena, V. Mishra, R. Kumar, P.R. Sagdeo, and Archna Sagdeo, Mater. Res. Express **2**, 96105 (2015).
[33] V. Džimbeg-Malčić, Ž. Barbarić-Mikočević, and K. Itrić, Tech. Gaz. **19**, 191 (2012).
[34] P. Kubelka, JOSA **38**, 448 (1948).
[35] P. Singh, I. Choudhuri, H.M. Rai, V. Mishra, R. Kumar, B. Pathak, A. Sagdeo, and P.R. Sagdeo, RSC Adv. **6**, 100230 (2016).
[36] E. Orhan, J.A. Varela, A. Zenatti, M.F.C. Gurgel, F.M. Pontes, E.R. Leite, E. Longo, P.S. Pizani, A. Beltràn, and J. Andrès, Phys. Rev. B **71**, 85113 (2005).
[37] W.J. Merz, Phys. Rev. **76**, 1221 (1949).
[38] Y.P. Varshni, Physica **34**, 149 (1967).
[39] J. Wu, W. Walukiewicz, W. Shan, K.M. Yu, J.W.A. Iii, S.X. Li, E.E. Haller, H. Lu, and W.J. Schaff, J. Appl. Phys. **94**, 4457 (2003).
[40] L.F. Jiang, W.Z. Shen, and Q.X. Guo, J. Appl. Phys. **106**, 13515 (2009).
[41] H. Hayashi, T. Nakamura, and T. Ebina, J. Phys. Chem. Solids **74**, 957 (2013).
[42] R. Naik, J.J. Nazarko, C.S. Flattery, U.D. Venkateswaran, V.M. Naik, M.S. Mohammed, G.W. Auner, J.V. Mantese, N.W. Schubring, A.L. Micheli, and A.B. Catalan, Phys. Rev. B **61**, 11367 (2000).
[43] F. Orapunt and S.K. O'Leary, Appl. Phys. Lett. **84**, 523 (2004).
[44] G.D. Cody, T. Tiedje, B. Abeles, B. Brooks, and Y. Goldstein, Phys. Rev. Lett. **47**, 1480 (1981).
[45] J.-H. Park and R.N. Blumenthal, J. Am. Ceram. Soc. **71**, C462 (1988).
[46] D.C. Johnston, Phys. Rev. B **74**, 184430 (2006).
[47] M. Meaudre, P. Jensen, and R. Meaudre, Philos. Mag. Part B **63**, 815 (1991).
[48] S.B. Roy, G.K. Perkins, M.K. Chattopadhyay, A.K. Nigam, K.J.S. Sokhey, P. Chaddah, A.D. Caplin, and L.F. Cohen, Phys. Rev. Lett. **92**, 147203 (2004).
[49] A. Shahee, D. Kumar, C. Shekhar, and N.P. Lalla, J. Phys. Condens. Matter **24**, 225405 (2012).
[50] R. Rawat, K. Mukherjee, K. Kumar, A. Banerjee, and P. Chaddah, J. Phys. Condens. Matter **19**, 256211 (2007).
[51] K. Binder, Rep. Prog. Phys. **50**, 783 (1987).

[52] K. Brajesh, K. Tanwar, M. Abebe, and R. Ranjan, Phys. Rev. B **92**, 224112 (2015).
[53] S.C. Bhargava, J.E. Knudsen, and S. Morup, J. Phys. C Solid State Phys. **12**, 2879 (1979).